\documentstyle[12pt]{article}
\begin{document}
\date{}
\bigskip
\hspace*{\fill}
\vbox{\baselineskip12pt \hbox{DTP 98-19}\hbox{hep-th/9804031}}
\bigskip\bigskip\bigskip

\begin{center}
{\Large {\bf AdS Membranes Wrapped on Surfaces of Arbitrary Genus}}

\vspace{48pt}

Roberto Emparan
\vspace{12pt}

{\sl Department of Mathematical Sciences}\\
{\sl University of Durham}\\
{\sl Durham DH1 3LE, UK}\\
{\it roberto.emparan@durham.ac.uk}

\vspace{72pt}

{\bf Abstract}
\end{center}
\begin{quote}{%
\small
We present and analyze solutions of $D=11$ supergravity describing the 
``near-horizon'' (i.e., asymptotically $AdS_4\times S^7$) geometry 
of M2-branes wrapped on surfaces of arbitrary genus.
We study the forces experienced by test M2-branes in such backgrounds,
and find evidence that extremal branes on surfaces of genera higher 
than the torus are unstable.
Using the holographic connection between $AdS$ spaces and superconformal
field theories in the large $N$ limit, we discuss the phases of the associated
$2+1$ dimensional theories. Finally, we also study the extension of these 
solutions to other branes, in particular to D2-branes.
}\end{quote}

\medskip

\newpage

\section{Introduction}
The solitonic solutions of supergravity theories \cite{stelle} have come
to play a 
fundamental role in our current understanding of M-theory. Very recently, 
it has been conjectured that by studying the region near the core 
of certain D- and 
M-branes one can extract information about the worldvolume 
dynamics of a large number of such parallel branes, i.e., about 
the dynamics of
superconformal field theories (SCFTs) in the 
large $N$ limit \cite{juan,gkp,wit1}. 
A rapidly 
growing number of recent papers is devoted to testing and extending this 
correspondence. In particular, those of direct relevance 
to the $2+1$ SCFT on the M2-brane include 
\cite{kkr,ckktp,imsy,gm,ccaff,aoy,min,hal,fkpz,gomis,wit2,halyo}.

In this paper we present new exact solutions of $D=11$ supergravity, which
can be interpreted as describing (in a sense to be explained below) the 
region near the horizon of M2-branes that wrap surfaces of arbitrary 
topology. First, we study the geometric features of the solutions. Next,
we follow \cite{wit1,wit2} in trying to obtain information about the 
phases of the (poorly understood) $2+1$ theory that describes the 
worldvolume dynamics of a large number of parallel M2-branes. Finally, 
we analyze the 
generalization of these solutions to other branes.

\section{Geometry of Anti-deSitter M2-branes}\label{geom}

Consider the long range supergravity fields describing flat, non-extremal 
M2-branes,
\begin{eqnarray}\label{flatm2}
ds^2 &=& H^{-2/3}(-{f dt^2} +dx_1^2 +dx_2^2) + 
H^{1/3}(f^{-1}dr^2 +r^2 d\Omega_7^2),\nonumber\\
F_{t x_1 x_2 r} &=& \sqrt{1+\left({r_0\over r_2}\right)^6} 
\partial_r \left({1\over H}\right), \\
H&=& 1 + \left({r_2\over r}\right)^6,\qquad f= 
1 - \left({r_0\over r}\right)^6 .
\nonumber
\end{eqnarray}
Here $r=r_0$ is the location of the
(outer) horizon. The geometry is asymptotic to flat 
Minkowski space at infinity. Now, consider the case where $r_0\ll r_2$, and
go to the region near 
the horizon. This amounts to setting
\begin{eqnarray}
H = \left({r_2\over r}\right)^6,\qquad 
F_{t x_1 x_2 r} &=& \partial_r \left({1\over H}\right) ,
\end{eqnarray}
while keeping $f$ and the rest of the solution as above. The resulting 
geometry
asymptotes to $AdS_4\times S^7$, the radius of $S^7$ being $r_2$. This
geometry can also be obtained by applying a series of U-duality and coordinate 
transformations to (\ref{flatm2}) \cite{hyun,sfet}, 
which suggests that both solutions  are somehow
equivalent in the full M-theory. In fact, recent developments have stressed the
fact that the dynamical aspects of branes are essentially encoded in the
region near the horizon.

Along the spatial worldvolume directions $(x_1, x_2)$ the horizon of this 
M2-brane solution is flat. But, in general, we can compactify $x_1$ and $x_2$
so that the brane wraps a 2-torus \footnote{Absence of singularities requires 
then that $t$ be compact too \cite{gib}.}. Our aim now is 
to construct solutions 
where the M2-brane wraps surfaces of other genera, like
2-spheres, or surfaces with two or more handles. We will always be
working in the region ``near the core,'' analogous to the $AdS$ region above. 
As will be apparent in a moment, these M2-brane 
solutions take, locally, the form of one of the following three families,
labelled by the parameter $\eta = +1, 0,-1$:
\begin{eqnarray}\label{topm2}
ds^2 &=& H^{-2/3}\left(-{f_\eta dt^2} +dx_1^2 + S^2_\eta(x_1)dx_2^2\right) 
+ H^{1/3}(f_\eta^{-1}dr^2 +r^2 d\Omega_7^2),\nonumber\\
F_{t x_1 x_2 r} &=& S_\eta(x_1) \partial_r \left({1\over H}\right), \\
H&=& \left({r_2\over r}\right)^6,\qquad f_\eta= 
1 +\eta \left({r_2\over r}\right)^4
- \left({r_0\over r}\right)^6,\nonumber
\end{eqnarray}
where
\begin{equation} 
S_\eta(x_1) =\left\{ \begin{array}{ll}
\sin\left( {2 x_1\over r_2}\right),& 
\eta = +1\\
 1, & \eta =0\\
\sinh\left( {2 x_1\over r_2}\right),&\eta = -1.
\end{array} \right.
\end{equation}
It is straightforward to check that these are solutions of the equations of
$D=11$ supergravity. Then,

\begin{itemize}

\item For $\eta =0$ we simply recover 
the near horizon solution of M2-branes wrapping 2-tori.

\item For $\eta = +1$, we see that $x_1$ and $x_2$, restricted to
$x_1 \in [0,\pi r_2/2)$, $x_2 \in [0,\pi r_2)$, parametrize a 2-sphere. 

\item For $\eta =-1$ we get the hyperbolic metric $H_2$ on 
the plane $(x_1, x_2)$. A standard result from the theory of
Riemann surfaces tells us that, by taking quotients of the 
universal covering of $H_2$ 
with discrete subgroups of its isometry group that act freely and properly 
discontinuously, we can generate the closed Riemann 
surfaces of any genus higher than $1$. 

\end{itemize}

These solutions are, therefore, enough to describe closed, compact, 
orientable M2-branes. Non-orientable surfaces, 
for any genus, can also be obtained by 
taking in addition quotients by discrete involutions. We will not be 
considering such branes explicitly, but much of what we will say 
is applicable to them, too.

The geometry of the solutions in (\ref{topm2}) splits again into the 
angular sphere $S^7$ of
constant radius, and the asymptotically $AdS_4$ spacetimes spanned by
$(t,x_1,x_2,r)$. In this sense, the four-form field strength has
produced a spontaneous compactification of the $D=11$ theory down to $D=4$
spacetime with a negative cosmological constant.
Interpreted this way, the four dimensional part of the solution corresponds 
to black holes with
horizons of arbitrary topology, which have been the focus of some interest 
recently 
(see \cite{mann} for extensive references).
 We will come back
to this connection in section \ref{holothermo}. For 
the moment, let us point out that the solutions with $r_0=0$ are all locally 
exactly $AdS_4$ (and not only asymptotically), differing from each other
only in identifications of points.

Let us analyze the existence of (outer) horizons in these solutions (compare
\cite{brill,vanzo}). 
If present, they
will correspond to the largest real roots of the equation $f_\eta=0$. For 
toroidal membranes 
$(\eta=0)$, these occur at $r=r_0$. As is well known, if $r_0=0$ then 
the zero is 
a double one and we obtain an extremal, supersymmetric, flat M2-brane. 

When $\eta = +1$, as long as $r_0\neq 0$ there exists a horizon, whose size
decreases with $r_0$. In the limit $r_0=0$
there is no horizon but the spacetime ($AdS_4\times S^7$)
is completely non-singular.

The situation for $\eta =-1$ is slightly
more complicated. It turns out that negative
values of $r_0^6$ (and therefore imaginary values of $r_0$) yield sensible
solutions (i.e., non-singular horizons) as long as they are bigger than 
a critical value,
\begin{equation} 
(r_0)^6 \geq (r_{0c})^6 \equiv -{2 (r_2)^6\over 3\sqrt{3}}.
\end{equation}
Then, when $r_0=r_{0c}$, the function $f_\eta$ has a double 
zero at $r=3^{1/4} r_2$.
Although this is not interpretable as a horizon, this extremal solution 
will be of
relevance later: it plays the role of the ground state for higher genus 
membranes. For larger values of $(r_0)^6$ we always find a 
nondegenerate horizon.

If we dimensionally reduce these solutions along the $x_2$ direction we obtain
string-like solutions of Type IIA supergravity. In particular, for $\eta =0$
we find the familiar solution near the core of a fundamental string. 
However, for 
$\eta \neq 0$ the dilaton in these solutions becomes singular
where $S_\eta (x_1) =0$.
E.g., for
$\eta =1$, reduction of a spherical M2-brane along its parallel circles
would yield an ``open
string'', and the singularities at $x_1=0,\pi$
would correspond to the endpoints of the 
string. Notice, however, that the singularity is present at every value of
$r$, and not just at the horizon. Thus, the interpretation as an ``open
string'' should not be taken too seriously.

{}Further reduction along the membrane worldvolume directions, say $x_1$, 
is hindered by the fact that $\partial/\partial x_1$ does not 
generate an isometry. For the
same reason, we can not directly apply a T-duality transformation 
along the ``open string'' solution. 

An important fact to notice is that, for genus different from $1$, the size 
of the membrane at the horizon is fixed once $r_2$ and $r_0$ are chosen. Using
the Gauss-Bonnet theorem, the area of the horizon for the genus $g$ membrane
$(g\neq 1)$ is
\begin{equation}
A_h = 4\pi |g-1| \left({r_h\over r_2}\right)^4,
\end{equation}
where $r_h$ is the value of $r$ at the (outer) horizon.
In contrast, for genus $1$, the size of 
the horizon of the toroidal membrane is arbitrary, since we 
can choose any periodicity for the
coordinates $x_1$ and $x_2$.

The charge density of these branes can be readily computed 
by integrating $*F$
over an angular $S^7$ at constant radius. In units where 
the eleven dimensional
Newton's constant is equal to $(16\pi)^{-1}$, the charge density is 
$q_2=2 \pi^4 (r_2)^6$. 
One would also like to have the energy density, or tension, of 
these branes. 
This is, however, more 
problematic. The ADM masses, or energy densities, are usually defined by 
integrals on a boundary in the asymptotically flat region. Evidently,
this we can not
do in the present case. There do exist definitions of mass 
in asymptotically anti-deSitter
spaces \cite{abdes}, but this is not enough. The definition
of mass in asymptotically $AdS_4$ space yields, 
for the flat M2-brane, only the energy density {\it above the extremal state}, 
since, in the $AdS$ spacetime
the extremal state is the natural choice for ground state\footnote{This energy
is, in fact the same as the thermodynamical energy that 
we will find in Sec.~\ref{holothermo}.}. 
In order to find the mass of the extremal state with respect to the
(Poincar\'e invariant) Minkowski vacuum, we would need to 
connect our solutions to that state by extending them 
to a suitable asymptotically flat region. At present we do not 
have any such extension. A shortcut might be provided by use 
of the BPS relation between tension and charge density, $T_2 = q_2$ (in 
the units chosen). The latter, however, need not be valid for 
M2-branes with topologies different
from the torus.

\section{Brane probes}\label{probes}

Parallel, flat,
extremal M2-branes do not exert any static force on each other. 
This is a consequence 
of their being BPS states, and it implies that we can stack an arbitrary 
number of these branes without any energy cost. Such systems are 
marginally stable.
It is of interest to see what are the forces between branes of 
other topologies.

An easy way to study the forces between parallel similar M2-branes
is by considering a light
(test) M2-brane in the background of a very massive assembly of similar, 
parallel M2-branes.
The test
M2-brane is described in terms of the action
\begin{equation}
I_{M2} =-T_2\int d^3\xi \sqrt{-\det g_{\alpha\beta}}
+{Q_2}\int A,
\end{equation}
where $g_{\alpha\beta}$ and $A_{\alpha\beta\gamma}$ are the pullbacks to the 
worldvolume of the spacetime metric and 3-form potential. Since we want 
to test the forces between M2-branes, we take (\ref{topm2}) as
backgrounds, and work in static 
gauge, where $\xi^\alpha = X^\alpha$. Note that, clearly, the test 
M2-brane has the same 
topology as those creating the background. 
The static interaction potential $V(r)$
that the test M2-brane experiences is obtained then 
from $I_{M2} = -\int dt V(r)$.
One easily finds
\begin{equation}
V(r) \sim H^{-1}(\sqrt{f_\eta} -1).
\end{equation}
It is a simple task to plot this potential for the different values of $\eta$,
extremal or non-extremal.
Probably, the most interesting case is that of extremal M2-branes. One finds
that the test M2-brane 

\begin{itemize}

\item  Is attracted by extremal spherical branes ($\eta=1, r_0=0$).

\item Experiences no static force in the background of 
extremal flat branes ($\eta=0=r_0$).

\item Is {\it repelled} by extremal higher genus branes 
($\eta=-1, r_0=r_{0c}$).

\end{itemize}

This can be taken as an indication that assemblies of spherical, toroidal,
and higher genus M2-branes are,
respectively, stable, marginal, and unstable. The instability of the solution
for M2-branes of higher genus is presumably present already at the classical
level, and is most likely due to the negative curvature of the worldsheet
of the brane.

If we go on to consider the force experienced by a test M2-brane in the 
background of non-extremal branes, we find that, at short enough distances,
the force is always attractive. As we increase $r$ and 
go to the asymptotically $AdS$ region, the behavior becomes the same as in the
extremal background. In particular, the test M2-brane in the non-extremal
$\eta =-1$ background is attracted near the horizon, but repelled at larger
distances.

\section{Thermodynamics and holography}\label{holothermo}

In \cite{wit1,wit2}, the thermodynamics of the $AdS_{p+2}$ 
part of the branes near the horizon has been used to extract 
information about
the phase structure of $(p+1)$-dimensional SCFTs
in the large $N$ limit. This correspondence is ``holographic'' in the sense
that the SCFT is associated to the 
boundary of the $AdS$ space.

Although one is ultimately
interested in the thermodynamics in the infinite volume limit, it was found
that there are some
interesting phenomena at finite volume. 
In \cite{wit1}, branes with the topology of spheres $S^p$, 
described by the Schwarzschild-Anti-deSitter
solution in $p+2$ dimensions, were considered. At low temperatures,
the dominant phase is described in terms of the no-black-hole, $AdS_{p+2}$ 
solution, which exhibits what could be called ``kinematic confinement.'' 
At higher temperatures, a phase 
transition takes place to the solution containing a black hole, which is 
interpreted as a deconfinement phase (see also \cite{klts}). 
High temperatures here correspond to large horizon sizes,
and the spherical brane is better and better approximated by a flat brane: 
finite size effects become unimportant. 
Thus, in the infinite volume limit the SCFT is described in terms of the
flat brane geometry \cite{wit2}.

In the present situation, we would be dealing with the large $N$ 
SCFT in $2+1$ dimensions associated to the worldvolume of a large 
number $N$ of parallel M2-branes.
This theory is, in fact, very poorly understood for any $N>1$. It is 
therefore very difficult to test the results
obtained from the holographic conjecture, which means that 
they should instead be taken as predictions. 

The solutions presented in Section \ref{geom} allow us
to discuss the theory on manifolds of arbitrary spatial topology. 
The phases associated to flat and spherical M2-branes are just like in
\cite{wit1}: spherical branes at low temperatures are described
by the manifold with $r_0=0$, but undergo a phase transition at higher 
temperatures to the manifold with $r_0\neq 0$. The theory on ${\bf R}^2$
(or on a large torus) is always in the high 
temperature phase. The entropy in this phase, however, scales as a puzzling
$N^{3/2}$, instead of the more usual ``deconfinement'' 
dependence on $N^2$ \cite{klts}.
We discuss now the phases of the $2+1$ theory 
on surfaces with more than one handle. 

Since, for the purposes of this Section, we do not need the 
$S^7$ part of the metric, use of 
the ``$D=11$ coordinates'' in (\ref{topm2}) becomes somewhat awkward, and it
may be convenient to use ``$D=4$ coordinates'' $(t,\theta,\varphi,\rho)$,
\begin{equation}
\rho={r^2\over 2 r_2},\quad \theta = {2 x_1\over r_2},\quad 
\varphi ={2 x_2\over r_2},
\end{equation}
and parameters $\ell = {r_2/ 2}$, $\mu = {r_0^6\over 4 r_2^5}$, in terms 
of which the 4-metric takes the form
\begin{eqnarray}\label{4dbh}
ds^2&=&-V_\eta dt^2 +V^{-1}_\eta d\rho^2 +\rho^2(d\theta^2 +S_\eta^2(\theta)
d\varphi^2),\\
V_\eta &=& \eta -{2 \mu\over \rho}+{\rho^2\over \ell^2}.\nonumber
\end{eqnarray}
In this form we can easily make connection with the recent discussion
on thermodynamics of $D=4$ topological black holes in \cite{brill,vanzo},
which we will reinterpret in the context of M-theory. In this parametrization,
the extremal solutions for $\eta=+1,0$ correspond to $\mu =0$, with horizon
at $\rho=0$, while for $\eta=-1$ they correspond to $\mu=\mu_c$ and horizon
at $\rho=\rho_c$, where
\begin{equation}
\mu_c=-{\ell\over 3\sqrt{3}}, \qquad \rho_c ={\ell \over \sqrt{3}}.
\end{equation}

A concept of the temperature of the horizon can be obtained from standard
Euclidean arguments, from the period of Euclidean time 
needed to avoid conical singularities at the horizon
\footnote{Since the solutions are not asymptotically 
flat, these 
will not be asymptotic temperatures. Nevertheless, they are still meaningful
quantities, see, e.g., \cite{brill}.}. 
The extremal solutions are zero 
temperature states (Euclidean time can be identified with arbitrary 
periodicity). For states above extremality
we obtain inverse temperatures $\beta$
\begin{equation}
\beta = {4\pi \rho_h \ell^2 \over 3\rho_h^2 +\eta \ell^2}=
{2\pi r_h^2 r_2^3 \over 3 r_h^4 +\eta r_2^4},
\end{equation}
where $\rho_h$ is the value of $\rho$ at the horizon. 
Notice that for large $\rho_h$ the temperature grows to a value independent of 
the topology of the horizon.

The partition function and free energy of the system are computed from the
classical action,
\begin{equation}
-\log Z =\beta F = I_{cl}-I_{cl}^0.
\end{equation}
The action of a reference background, $I_{cl}^0$, must be subtracted for 
regularization. This reference background acts as a ground 
state. For flat and spherical branes the choice is clear: it is the 
solution with $r_0=0$, which is locally $AdS_4$, and can be at 
zero temperature. 
However, for higher genera ($\eta =-1$) the
locally $AdS_4$ solution ($r_0=0=\mu$) is not the same as 
the extremal, zero temperature one, $\mu=\mu_c<0$. 
Now, computation of the
action requires matching the Euclidean geometries of the excited 
and the ground state at the 
asymptotic boundary. It does not seem adequate to take $AdS_4$ 
as the reference background. The reason is that
absence of conical singularities at the horizon in the solution 
with $\mu=0$ requires a specific value of Euclidean time periodicity
$\beta$. Matching the boundary geometry of this solution to one with
$\mu\neq 0$ will introduce singularities at the horizon. In contrast, 
in the extremal solution with $r_0=r_{0c}$ one can identify Euclidean
time with 
arbitrary period without introducing singularities. This suggests 
that it is the correct reference state
\footnote{This is the view advocated
in \cite{vanzo}. It also eliminates the notion of negative mass black holes
in the $D=4$ context, see eq.~(\ref{thermen}) below.}. It is somewhat 
worrying, nevertheless, that, as we argued in 
Sec.~\ref{probes}, this
ground state appears to be an unstable one. 

Having taken the $\mu=\mu_c$ state as the reference background, 
the calculation of the
action is straightforward (see, e.g., \cite{wit2}, or \cite{vanzo}), 
and one finds (hereafter we set $\eta=-1$)
\begin{equation}\label{action}
\beta F =-{\Omega_g\over 4}{\rho_h^4 +\rho_h^2 \ell^2\over 3\rho_h^2 -\ell^2}
-\beta \mu_c {\Omega_g\over 4\pi}.
\end{equation}
Here $\Omega_g=4\pi(g-1)$ is the volume of the unit surface of genus $g$. 
We can compute now the thermodynamic energy
\begin{equation}\label{thermen}
E = {\partial(\beta F)\over \partial \beta} = 
{\Omega_g\over 4\pi}(\mu -\mu_c).
\end{equation}
This is always positive.
Usually, in terms of flat M2-branes, this energy corresponds to the mass 
above the mass of the extremal state. 
This interpretation may also be adequate 
for M2-branes with different topologies. The thermodynamic
entropy is given by
\begin{equation}
S=\beta (E-F) = {\Omega_g\over 4}\rho_h^2 = {A_h\over 4},
\end{equation}
the usual Bekenstein-Hawking formula. Had we chosen the $AdS_4$ solution as
background we would have found a different result, as well as negative
values for $E$ \cite{vanzo}. Notice that the ground state has non vanishing
entropy, in contrast with flat and spherical branes; we do not have any good
explanation for this fact. Finally, we 
find the specific heat 
\begin{equation}
C= -\beta{\partial S\over \partial\beta} =
{\Omega_g \rho_h^3\over 2}{3\rho_h^2 -\ell^2\over 3\rho_h^2 +\ell^2}.
\end{equation}
This is always positive. What we learn from here is that the 2+1 SCFT 
defined on a surface of genus higher than $1$ does not have 
phase transitions at finite volume. 
There is no
``kinematic confinement'' at low temperatures, and the theory is always in
the ``high temperature'' phase. Also, it should be clear that,
as we go to the infinite volume limit,
the higher genus manifold becomes indistinguishable from the flat manifold.

\section{Other branes}

A natural question to ask is whether one can find similar solutions for
other branes. The most natural case to consider is that of a D2-brane 
on an arbitrary genus surface $\Sigma_g$. But one could also think of 
$p$-branes, $p>2$, on, say, $\Sigma_g\times T^{p-2}$. For simplicity we
give details 
only for D2-branes, the generalization to other situations
presenting no further novelties.

We will only consider the cases with parameter $r_0=0$. For our 
present purposes, there are two main differences between the 
supergravity solutions corresponding to the D2-brane and the M2-brane. 
First, the D2-brane has non-trivial dilaton.
Second, the metric of the flat D2-brane near the horizon
does {\it not} split into $AdS_4\times S^6$ neither in
Einstein frame $ds^2_E$ nor in string frame, $ds^2_S$. 
However, it does so in a conformally related frame,
\begin{eqnarray}\label{d2}
d\bar s^2 &=& e^{-2\phi/5}ds^2_S=
e^{\phi/10}ds^2_E\nonumber\\
&=& H^{-3/5}(-{dt^2} +dx_1^2 +dx_2^2) + 
H^{2/5}({dr^2} +r^2 d\Omega_6^2),\\
H&=&\left( {r_2\over r}\right)^5,\qquad e^{\phi}=H^{1/4}.\nonumber
\end{eqnarray}
That the $(t,x_1,x_2,r)$ part of this metric is locally isometric 
to $AdS_4$ can be easily seen by changing to ``$D=4$ coordinates,''
\begin{equation}\label{10d4d}
r=\left({9 r_2 \rho^2\over 4}\right)^{1/3},\quad x_1=\ell \theta,\quad
x_2 = \ell\varphi,\quad \ell\equiv{2r_2\over 3},
\end{equation}
in terms of which the 4-metric  becomes
\begin{equation}\label{4dtorbh}
d\bar s^2_{(4)}=-{\rho^2\over \ell^2} dt^2 +{\ell^2\over \rho^2} 
d\rho^2 +\rho^2(d\theta^2 +d\varphi^2),
\end{equation}
i.e., the metric (\ref{4dbh}) with $\eta =0=\mu$. 
We know that the solutions (\ref{4dbh}) with $\mu=0$ are, for all three values
of $\eta$, locally identical to $AdS_4$. 
Therefore, the solutions with $\eta =\pm 1$, $\mu=0$,
which correspond to different topologies, must be locally
related to (\ref{4dtorbh}) by a change of coordinates (globally
one must change the identification of points). Explicitly, if in 
(\ref{4dtorbh}) we change
\begin{eqnarray}\label{transf}
t&\rightarrow&\ell{\sqrt{\rho^2+\eta\ell^2}\sin(\sqrt{\eta}t/\ell) 
\over \sqrt{\eta}\Delta},\nonumber\\
\theta&\rightarrow&{\rho\sin(\sqrt{\eta}\theta)\cos\varphi 
\over \sqrt{\eta}\Delta},\nonumber\\
\varphi&\rightarrow&{\rho\sin(\sqrt{\eta}\theta)\sin\varphi 
\over \sqrt{\eta}\Delta},\\
\rho&\rightarrow& \Delta,\nonumber\\
\Delta&\equiv&\rho\cos(\sqrt{\eta}\theta) + \sqrt{\rho^2 +\eta\ell^2}
\cos(\sqrt{\eta}t/\ell),\nonumber 
\end{eqnarray}
for $\eta=\pm 1$, then we find the desired forms of the metrics,
\begin{eqnarray}\label{4dtopbh}
d\bar s^2_{(4)}\rightarrow d\bar s^2_{(4)}&=&-\left(\eta+
{\rho^2\over \ell^2}\right) dt^2 +
{d\rho^2\over \eta +\rho^2/ \ell^2} +\rho^2\left(d\theta^2 +
S^2_\eta(\theta)d\varphi^2\right),\\
S_\eta(\theta) &=& \sqrt{\eta} \sin(\sqrt{\eta}\theta).\nonumber 
\end{eqnarray}

Now it should be clear how to proceed to find D2-branes with topologies 
different from the torus: use (\ref{10d4d}) to express the transformation 
(\ref{transf}) in terms of ``$D=10$ coordinates,'' and then apply it
to the solution (\ref{d2}). The resulting metric can be conformally 
rescaled to string (or Einstein)
frame using the (transformed) dilaton, to find
\begin{eqnarray}
ds^2_S =e^{\phi/2}ds^2_E&=& A_\eta^{-1/2}\left[ H^{-1/2}\left(-{f_\eta dt^2} 
+dx_1^2 + 
S^2_\eta(x_1)dx_2^2\right) \right.\nonumber\\
&&+ 
\left. H^{1/2}(f_\eta^{-1}dr^2 +r^2 d\Omega_6^2)\right],\nonumber\\
e^{\phi} &=&H^{1/4} A_\eta^{-5/4}, \\
H&=& \left({r_2\over r}\right)^5,\qquad f_\eta= 
1 +\eta \left({r_2\over r}\right)^3,\nonumber
\end{eqnarray}
where, now
\begin{equation} 
S_\eta(x_1) =\left\{ \begin{array}{ll}
\sin\left( {3 x_1\over 2 r_2}\right),& 
\eta = +1\\
 1, & \eta =0\\
\sinh\left( {3 x_1\over 2 r_2}\right),&\eta = -1,
\end{array} \right.
\end{equation}
and
\begin{equation} 
A_\eta(t,x_1,r) =\left\{ \begin{array}{ll}
\cos\left( {3 x_1\over 2 r_2}\right)+f_1^{1/2} 
\cos\left( {3 t\over 2 r_2}\right),& 
\eta = +1\\
 1, & \eta =0\\
\cosh\left( {3 x_1\over 2 r_2}\right)+f_{-1}^{1/2} 
\cosh\left( {3 t\over 2 r_2}\right),&\eta = -1.
\end{array} \right.
\end{equation}

It is in the conformal factor, or equivalently in the dilaton, where
the complication resides: for $\eta=0$ the dilaton depends only on $r$, 
but for $\eta=\pm 1$ it acquires also a
dependence on $x_1$ and, worse, on $t$, which seems an undesirable 
feature of these solutions. Notice that the time dependence vanishes at
the null Killing surface (which is a horizon for $\eta =-1$) 
where $f_\eta=0$.
However, for the spherical brane ($\eta=+1$), the conformal factor 
becomes singular at the
equator ($x_1 = \pi r_2/3$) of the horizon.

This procedure is not valid for solutions with $r_0\neq 0$, since these are
locally different from each other. However, one would expect the solutions
for these cases to present a similar dependence of the dilaton on $t$ and
$x_1$. 

In the case of higher $p$-branes, if we start from the flat, extremal
solution near the core, compactification on $T^{p-2}$ will yield
a metric that, again, can be conformally rescaled to look like 
$AdS_4\times S^q$. Then we can repeat the steps above to find spatial
worldvolumes of the form $\Sigma_g\times T^{p-2}$.

However, the time dependence and singularities of the dilaton obscure 
the significance 
of these solutions for dilatonic branes with non-standard topologies.

\section{Final remarks}

Usually, one wants to view extremal $p$-branes as solitons interpolating
between two different vacua of supergravity: the one near the core 
($AdS_{p+2}\times S^q$),
and the Minkowski vacuum, $H=1=f$ in (\ref{flatm2}) \cite{gibtow}. 
In the present situation, for non-toroidal M2-branes, 
one should note that, first, the solution near the core is not 
supersymmetric. Second, the fields that result from setting
$H=1=f_\eta$ in (\ref{topm2}), with $S_\eta(x_1)\neq 1$ 
do {\it not} solve the equations of $D=11$ supergravity. It is not clear
at present what, if any, is the generalization of (\ref{topm2}) with 
suitable asymptotics (flat in directions transverse to the spatial
directions of the brane), in a way similar to the asymptotically flat 
solution (\ref{flatm2}).

Closely related to this, there is one obvious shortcoming in 
the M2-brane solutions that we have introduced:
we have no way to define the total energy density of the M2-brane (i.e., 
the energy measured with respect to a vacuum that is Poincar\'e invariant in
directions transverse to the M2-brane spatial directions), only the 
energy above the extremal state. Indeed,
the fact that extremal M2-branes of higher genera seem to repel each other
may be an indication that, for them, the energy density might be much larger
than for flat branes.
Certainly, they are not BPS states, so we should not, presumably, 
expect their energy density to be fixed by the value of their charge.

Other recent works have considered
modifications of the basic, flat M2-brane solution near the core 
that change the geometry
{\it transverse} to the brane \cite{halyo}, i.e., substituting the
angular $S^7$ with a different manifold. Here, in contrast, 
we have modified the geometry of the brane itself. Both sorts of modifications 
can be considered and applied independently.

\section*{Acknowledgements}
We would like to thank Ivo Sachs for conversations, and Igor Klebanov 
and Jeremy Michelson for comments on a previous version of the manuscript.
This work has been partially 
supported by EPSRC through grant GR/L38158 (UK), and by grant 
UPV 063.310-EB225/95 (Spain).

\end{document}